\documentstyle[preprint,aps]{revtex} % for normal preprin style
\def\x{\times}
\begin{document}
\draft
\preprint{}
\title{Effects of Gravitational Smearing on Predictions \\ of Supergravity
 Grand Unification }
\author{T.Dasgupta, P.Mamales and Pran Nath}
\address{
Department of Physics,
Northeastern University,
Boston,  MA 02115}
\date{\today}
\maketitle
\begin{abstract}
 Limits on gravitational smearing from the dominant
dimension 5 operator(with strength characterised by $cM/2M_{Planck}$)
are obtained using the
LEP data in supersymmetric SU(5) grand unification.Effects of c
 on the quasi infrared fixed point solutions of $m_b/m_{\tau}$
unification are also analysed.It is found that $c>0$ softens and $c<0$
stiffens the fixed point
constraint.
The effect of c on the upper bound of the Higgs triplet mass is analysed
and the existence of a scaling induced by c is discussed.
Tests for the
existence of gravitational smearing
are discussed.
\end{abstract}

\narrowtext
%\begin{multicols}{2}
The high precesion LEP data on the gauge coupling constants
measured at the $Z$--scale [1] appear encouraging [2] for
ideas of
supersymmetry and grand unification and has spawned considerable activity
towards extraction of further predictions from supersymmetric grand unification
[3,4].
However it has been pointed out that gravitational smearing
[5,6] from the unknown Planck scale physics may affect predictions of
supersymmetric grand unification [7].  The analysis of Ref[7] ,however,
 was limited in that it used one loop renormalization group
evolution, ignored the top mass dependence and did not include the constraints
of $b/\tau$ unification.
  In this Letter we use LEP data to determine the allowed range of
gravitational smearing parameter c using the full 2 loop renormalization
group (R.G.) analysis of gauge  coupling constants as well as  two loop
evolution of $t,b,\tau$ yukawas.
We find that current data already puts important constraints on gravitational
smearing. Simultaneously we analyse  the
effect of c on the quasi infrared fixed point solutions in the top quark yukawa
coupling under the constraint that $b/\tau$ yukawas unify at the
 GUT scale[8,9,10].
The dependence of the upper bound on c is investigated and the phenomenon that
the quantum gravity corrections generate an effective scaling of the heavy
thresholds is discussed.
We also emphasize that the gaugino sector involves an additional model
dependent
parameter  $c'$ which has the same origin as c but is in
general numerically different.We discuss tests for determinations of c and
$c'$.

The framework of the analysis we present here is the supergravity SU(5) model
[11,12]. We assume that the GUT symmetry  of this model is broken by
the term $\lambda_1[\frac{1}{3} \Sigma^3 + \frac{1}{2} M\Sigma^2]$ while the
Higgs triplet becomes superheavy and the Higgs doublets remain light via
the interaction $ \lambda_2 H_2 [\Sigma + 2M']H_1[13]$,
where $H_1 ( H_2 )$ are $ 5 (\bar{5})$ of Higgs and $\Sigma$ is a $24$--plet of
SU(5).  The superheavy fields after spontaneous breaking of the GUT symmetry
consist of $(3, 2, 5/3) + (\bar{3}, 2, -5/3)$ massive vector bosons of mass
$M_V = 5\sqrt{2}gM$, $(1,3,0) + (1,\bar{3},0)$ massive color Higgs triplets of
mass  $M_{H3} = 5 \lambda_2 M$, $(1,8,0)+(1,3,0)$ massive
$\Sigma$--fields of mass $M_{\Sigma}=5 \lambda_1 M/2$ and a singlet $\Sigma$
field
of mass  $M_{\Sigma}/5$, where M enters in the VeV of $\Sigma$ as
$<diag(\Sigma)>=M(2,2,2,-3,-3)$.
In our analysis we use 2--loop evolution of gauge coupling constants which are
given by
[14]
\begin{mathletters}
\begin{equation}
 \frac{d\gamma_i^{-1}}{\!\!\!\!\!\!\!\!dt}=
\left[ b_i + \sum_{j=1}^3 b_{ij}\gamma_j
- \sum_{j=t,\beta,\tau} a_{ij}Y_j \right] \label{RGE3}
\end{equation}
where $\gamma_i = \alpha_i/4\pi \  (\alpha_i=g_i^2/4\pi)$,
$ Y_i = \lambda_i^2/16\pi^2$,
t=2log($M_G/Q$) and the $b_i$ are given by
\begin{eqnarray}
b_1 = &&\frac{33}{5} +b_{1lt}
- 10\Theta_{\bar{V}} + \frac{2}{5}\Theta_{\bar{H}_3}\\
b_2 = && 1+ b_{2lt}
+ 2\Theta_{\bar{\Sigma}} -6\Theta_{\bar{V}}\\
b_3 = && -3+ b_{3lt}
+ 3\Theta_{\bar{\Sigma}} - 4\Theta_{\bar{V}} + \Theta_{\bar{H}_3}
\end{eqnarray}
\end{mathletters}
where $b_{ilt}$ are the corrections from the light thresholds,
and $\Theta_{\bar{V}}\equiv \Theta(\mu - M_V)$etc.
We use the standard match and run technique
between different thresholds and in going from   $\bar{DR}$ to $\bar{MS}$
as we cross the highest SUSY threshold we
 use the correction $\Delta_{ic} =
C_2(G_i)/12\pi$ where $C_2(G_i)$ is the quadratic Casimir.

We discuss next the influence of Planck scale physics
on GUT analyses. These effects manifest by the
appearance of higher$(>4)$ dimension operators at the scale of GUT physics.
For example in N=1 supergravity the kinetic energy and mass terms for the
gauge fields and the gaugino are given by [12]
%\end{multicols}
\widetext
\begin{eqnarray}
e^{-1}{\cal L}=&& -\frac{1}{4}\Re\!\left[f_{\alpha\beta}F_{\mu\nu}^{\alpha}
F^{\beta\mu\nu}\right] + \frac{1}{4} i\Im\!\left[
f_{\alpha\beta}F_{\mu\nu}^{\alpha}\tilde{F}^{\beta\mu\nu}\right]
+ \frac{1}{2}\Re\!\left[ f_{\alpha\beta}\left(
- \frac{1}{2}\bar{\lambda}^{\alpha}D\!\!\!\!/\lambda^{\beta}\right) \right]
\nonumber \\
& & - \frac{1}{8} i \Im\!\left[ f_{\alpha\beta} e^{-1}D_{\mu}
(e \bar{\lambda}^{\alpha}\gamma^{\mu}\gamma_5\lambda^{\beta})\right]
+ \frac{1}{4}\bar{e}^{G/2}G^a(G^{-1})^b_a(\partial f^*_{\alpha\beta}
/\partial z^{*b}\lambda^{\alpha}\lambda^{\beta}) + {\rm h.c.}
\end{eqnarray}
%\begin{multicols}{2}
\narrowtext
where $z^a$ are the scalar fields $\lambda^{\alpha}$ are the gauginos
and the function $G$ is given by $G = -\ln[\kappa^6 W W^*]- \kappa^2 d$, where
$W$ is the superpotential, $d(z,z^*)$ is the Kahler potential and $M_P$=
$\kappa^{-1} = (8\pi G_N)^{1/2} = 0.41\times 10^{-18}$ GeV where $G_N$ is
Newton's
constant.  Now in general
$f_{\alpha\beta}$  can have a non-trivial field dependence and
one may write $f_{\alpha\beta} = A\delta_{\alpha\beta}+Bd_{\alpha\beta\gamma}
\Sigma^{\gamma}$, where
$d_{\alpha\beta\gamma}=2{\rm tr}(\{ \lambda_{\alpha}/2, \lambda_{\beta}/2\}
\lambda_{\gamma}/2)$ and $\lambda_{\alpha}/2$ are matrices in the adjoint
representation of SU(5) group generators, and $\Sigma^{\gamma}$ is defined by
$\Sigma^{a}_{b}= \Sigma^{\alpha} (\lambda_{\alpha}/\sqrt 2)^{a}_b$.
After spontaneous breaking of supersymmety via the hidden sector and the
breaking of
SU(5) to $SU(2)_L\x U(1)\x SU(3)_C$ from the VeV growth of $\Sigma$
 ,one can carry out a rescaling to achieve a
canonical kinetic energy for the gauge fields and the gaugino fields.One
finds then a splitting of the gauge coupling constant at the GUT scale $M_G$
(taken to be the highest mass threshold) so that
$\alpha_i^{-1}(M_G)=\alpha_G^{-1}(M_G) + c\epsilon_i$
where $\epsilon_i=n_i(\alpha_G)^{-1}M/(2M_P)$ and $n_i=(-1,-3,2)$ for i in
the $U(1)\x SU(2)_L\x SU(3)_C$ sectors.
After rescaling the gaugino mass matrix is
\begin{equation}
m_{\alpha\beta}= \frac{1}{4}\bar{e}^{G/2}G^a(G^{-1})^b_a
(\partial f^*_{\alpha\gamma}
/\partial z^{*b})f^{-1}_{\gamma\beta}
\end{equation}
Eq(3) can be reduced further.Thus
 writing $G^b(G^{-1})^ab=q_A D^{Aa}$ (A=1,2) where $D^{1a}=\delta^{az}$,
 $D^{2a}=\delta^{a\alpha} \Sigma^{*\alpha}_0$,defining $\vec{u}=(1,q_1,q_2)$,
$\vec{v}=(2\Sigma^2_0B'_{\Sigma},B'_z,B)$ and $ \vec{w}=(0,A'_z,
2\Sigma^2_0A'_{\Sigma})$ where $A'_z=
(\partial A/\partial z)_0, A'_{\Sigma}=(\partial A/\partial \Sigma^2)_0$,
and $B'_
z$ and $B'_{\Sigma}$ are
similarly defined, one finds that the gaugino masses are given by
\begin{equation}
\frac{M_i}{m_{1/2}} = \frac{\alpha_i}{\alpha_G} \left( 1 +c'
\frac{M}
{M_P}
n_i \right) \label{GauginoMass}
\end{equation}
where  $c' \equiv ( u.v/u.w)$, in accord with the analysis of Ref[6].
We observe then that a new parameter $c'$ enters the
neutralino and chargino masses. Thus, for
instance,the scaling mass relations which are known to hold over the
a majority of the parameter space allowed by the radiative electro-weak
symmetry breaking[15]
are now modified because of eq(4) and one has
\begin{equation}
(1+\Delta_1)m_{\tilde{W}_1}\simeq  2(1+\Delta_2)m_{\tilde{Z}_1},m_{\tilde{W}_1}
\simeq m_{\tilde{Z}_2}
\end{equation}
where $\Delta_1=-3c'M/M_P$ and $\Delta_2=\Delta_1/3$.Thus while the
$m_{\tilde{W}_11}-m_{\tilde{Z}_2}$
scaling relation is unchanged, there
can be a significant modification of the $m_{\tilde{W}_1}-m_{\tilde{Z}_1}$
scaling law depending on the value of $c'$.This modification will reveal
the existence of $c'$ and provide a measurement of it.
We note that the effects of $c$ in the R.G. analysis of gauge couplings
appear on an equal
footing to those of heavy thresholds.
Thus the coupling constant evolution in the neighborhood of the heavy
thresholds
can be written as $\alpha_i^{-1}(Q)= \alpha_G^{-1} + C_{ia}\log(M_a/Q)$ where
$M_a$ are the heavy thresholds and $C_{ia}$ can be read off from eq(1) and
$\alpha_G$ is evaluted at $M_G$ ( which we take to be the $M_{H_3}$ mass).
Inclusion of quantum
gravity effects modify this relation so that
\begin{equation}
\alpha_i^{-1}(Q)=\alpha_G^{-1} + \frac{cM}{2M_P}\alpha_G^{-1}n_i
+ C_{ia}\log\frac{M_a}{Q}
\end{equation}
Now by a transformation $M_a=M_a^{eff}e^{\chi_a}$
 one can absorb
 the quantum gravity correction by defining effective heavy thresholds so that
$\alpha_i^{-1}(Q)={\alpha^{eff}_G}^{-1}+C_{ia}\log(M_a^{eff}/Q)$.
Here  $\alpha^{eff}_G$
is $\alpha_G$ evaluated at $M^{eff}_G$ where  $M^{eff}_G$=$M_G$
exp(-5$C_P$),$C_P=(\frac{\pi cM}{M_P}\alpha_G^{-1})$  so that
 $(\alpha^{eff}_G)^{-1}=\alpha_G^{-1}-(15/2\pi)C_P$.Thus c generates
an effective scaling of the heavy masses which are described by
\begin{equation}
M^{eff}_a=M_a e^{-k_a C_P};k_a=(-\frac{3}{5},\frac{3}{10},5)
\end{equation}
where a=1,2,3 refer to $\Sigma,V,M_{H_3}$ masses.
We shall refer to this scaling again
when we discuss the numerical analysis of the R.G. equations.
Thus at this level of the R.G. analysis
quantum gravity effects can be masked by threshold
corrections.The situation here is similar to the correspondence between the
effects of the light thresholds and heavy thresholds [17].
We note that the effect of the $c'$ term is not masked by threshold corrections
as may be seen in eqs(4) and (5).
The confusion between threshold effects and quatum gravity effects can also
be removed if a $p\rightarrow\pi^0e^+$ or a $p\rightarrow\bar{\nu}K^+$
decay mode was observed.Thus the $p\rightarrow \pi^0e^+$ mode will determine
$M_V$ which combined with the determination of ${M^{eff}_V}$ from R.G. analysis
as discussed above gives on using eq(7)
\begin{equation}
 c=\frac{100}{3}\sqrt{\frac{2}{\pi}}\alpha_G^{3/2}\frac{M_P}{M_V}\log
\frac{M_V}{M_V^{eff}}
\end{equation}

Eq(8) provides a clean  determination of $c$ since $c'$ decouples in
this mode.
A similar relation holds for $M_{H3}^{eff}$ and $M_{H3}$
as can be read off from eq(7).  However,here the
extraction of
$M_{H3}$ from $p\rightarrow\bar{\nu}K$ mode depends also on $c'$ which enters
in
 the
chargino mass on which the  $p\rightarrow\bar{\nu}K$ decay mode depends.
Thus a knowledge of the neutralino and chargino spectrum will be
needed along with the p-decay data to fix both c and $c'$.

We discuss next the $m_b/m_\tau$ mass ratio.
We have carried out this analysis on a seven dimensional space parametrized
by $\alpha_{G}$, $M_{\Sigma}$,$M_{V}$, $M_{H_3}$, $M_{\rm susy}$,$\tan\beta$,
and c, where we have parametrized the low energy scales by one common scale
 $M_{\rm susy}$. We use two-loop
evolution between $M_{\rm susy}$ and the GUT scale.
These solutions are then matched on to the solutions below the SUSY
scale using the boundry condition
$\lambda_t(M_{SUSY}^-)=\lambda_t(M_{SUSY}^+)\sin\beta$, and
$\lambda_i(M_{SUSY}^-)=\lambda_i(M_{SUSY}^+)\cos\beta
(i = b,\tau)$.   We also take into account the GUT threshold
corrections in $b/\tau$ evolution [16].
In our analysis we used
$\alpha_{em}^{-1}(M_Z) = 127.9\pm 0.1$ and
included an $M_t$ dependence in $\sin^2\theta_w$ via the equation [10]
$\sin^2\theta_w = 0.2324 \pm 0.0003 +\delta$ where
$\delta=- 0.92\times 10^7 {\rm GeV}^{-2} \Delta^2$ and
$\Delta^2=[(M_t^{\rm pole})^2 - (143{\rm GeV})^2]$.
The range of acceptable values  of $\alpha_3$ are taken to be $0.12\pm 0.01$.
There are also constraints on $M_{\Sigma}$ and $M_{H_3}$ from perturbativity
of the GUT Yukawa couplings which we assume to imply $\lambda_{1,2}^2/4\pi
\leq 1/2$.
We have used a value of $M_t = 174 \pm 16$GeV as indicated by the CDF data
[18].   The $\pm 16$GeV variation of the top mass has significant effect
on the analysis  this dependence enters importantly via. R.G. effects
%and via. the $\Delta^2$ dependence.

In the determination of the $b/\tau$ masses we used renormalization group
evolution up to three--loop order in QCD and one--loop order in QED.
The evolution
from QCD is given by [8,19].
%\end{multicols}
\widetext
\begin{eqnarray}
%------------The QCD Running(3-loop) for quark mass----------------
m_f(\mu) & = & \hat{m_f} \left( -\beta_{1}\frac{\alpha_{s}(\mu)}{\pi} \right)
^{-\gamma_{1} / \beta_{1}}
\left\{ 1+\frac{\beta_{2}}{\beta_{1}} \left( \frac{\gamma_{1}}{\beta_{1}}
-\frac{\gamma_{2}}{\beta_{2}} \right) \frac{\alpha_{s}(\mu)}{\pi}+\frac{1}{2}
\left[ \frac{\beta_{2}^{2}}{\beta_{1}^{2}} \left( \frac{\gamma_{1}}{\beta_{1}}
-\frac{\gamma_{2}}{\beta_{2}} \right) ^{2} \right. \right.
\nonumber \\
& & \left. \left.
-\frac{\beta_{2}^{2}}{\beta_{1}^{2}} \left( \frac{\gamma_{1}}{\beta_{1}}-
\frac{\gamma_{2}}{\beta_{2}} \right) +\frac{\beta_{3}}{\beta_{1}}
\left( \frac{\gamma_{1}}{\beta_{1}}-\frac{\gamma_{3}}{\beta_{3}} \right)
\right]
\left( \frac{\alpha_{s}(\mu)}{\pi} \right) ^{2} \right\}
\end{eqnarray}
%\begin{multicols}{2}
\narrowtext
where $\beta_i, \gamma_i (i = 1,2,3)$ are given by
$\gamma_1=2$,
$\gamma_2 = \frac{101}{12}-\frac{5}{18}n_f$,
$\beta_1=-\frac{11}{2}+\frac{1}{3}n_f$,
$\beta_2 = -\frac{51}{4}+\frac{19}{12}n_f$
and $\beta_3$ and $\gamma_3$ are given by
\begin{eqnarray*}
\gamma_3&=&\frac{1}{96}\left[3747-(160\zeta(3)+\frac{2216}{9}n_f-\frac{140}{27}
n_f^2\right] \nonumber \\
\beta_3&=&\frac{1}{64}\left[-2857+\frac{5033}{9}n_f-\frac{325}{27}n_f^2\right]
\end{eqnarray*}
and the effect of QED corrections are determined by ,
$m_{f}(\mu)  =  m_{f}(\mu')\left(\frac{\alpha(\mu)}{\alpha(\mu')}\right)
^{\gamma_{0}^{QED}/b_{0}^{QED}}
$where
$b_{0}^{QED}  =  \frac{4}{3}(3\sum Q_{u}^{2}+3\sum Q_{d}^{2}+\sum Q_{e}^{2})
, \;
\gamma_{0}^{QED}  =  -3Q_{f}^{2}$ and
\begin{eqnarray*}
%-------the rel bet. ren. coupling and fine str. comnst.------------------
\alpha^{-1}(\mu) & = & \alpha_{em}^{-1}+\frac{1}{6\pi}
-\frac{2}{3\pi}\sum_{f}Q_{f}^{2}
\ln\frac{\mu}{m_{f}}\theta(\mu-m_{f})
\end{eqnarray*}
Unlike the analysis of [8], the value of $\alpha_3$
was determined self consistantly using eqs.(1) and (9).

We discuss now the major results of this paper. In Fig.1 we give a
determination of the range of c  using the LEP data on $\alpha_i$ under
the combined constraints of gauge coupling with and without $b/\tau$
unification,
when $M_{SUSY}$ lies in the range $M_Z-10TeV$,$M_{H_3}\geq 1\times 10^{16}$ and
all other parameters
are integrated out.
One finds that values of $c$ consistent
with the $1 \sigma $  LEP bound on $\alpha_3$ lie in the range
$-1.9 \leq c\leq 3.0$($-0.4 \leq c\leq 3.0$) with(without) $b/\tau$
unification. Thus remarkably the current LEP data already puts rather
stringent bounds on $c$.
Fig 1 also exhibits the fact that $b/\tau$ unification gives an upper bound on
$\alpha_3$
of $\alpha_3\leq 0.12$ over the allowed domain of c.
This upper limit holds for the $m_b$ mass range
$m_b\leq 4.8$ GeV.Fig.~2 shows the  upper bound on $M_{H_3}$ mass for various
values of c. One finds that the upper bound  depends sensitively on c
and exhibits a scaling in c as anticipated in our discussion
earlier ( see the discussion following eq(6)).
Fig.~2 shows that  a $c>0$ requires  a larger $M_{H_3}$ to generate the
same $M^{eff}_{H_3}$ as for c=0. This is what the scaling relation of eq(7)
implies.The magnification factor
%of the upper limits
$M_{H_3}(c=1)/M_{H_3}(c=0)$ is also consistent with what eq(7) gives.
We have  analysed the effect of
gravitational smearing on the infrared fixed point solutions of $M_t-tan\beta$
that arise when one imposes $b/\tau$ unification.Results are exhibited in
Fig.3 for several
values of c and several values of $\alpha_3$. For c=0 we find the standard
result that $b/\tau$ unification
puts $m_t$ close to its fixed point value. Positive values of c are found to
 help relax the fixed point constraints while negative
values of c  make the fixed point constraint even more
stringent.These features persist for the range of b quark masses
discussed above.

In conclusion we have discussed several phenomena due to the effect of
gravitational smearing on
analyses of supersymmetric SU(5) unification.
% and
%$b/\tau$ unification constraints.
 Remarkably one finds that the current
LEP data
already restricts c to lie in a narrow range.
We also discussed the effects on the constraints imposed by the infrared
fixed point in the top yukawa coupling and analysed the effects of c on
$M_{H_3}$.The existence of a scaling of GUT masses induced by c was
discussed.
 The analysis gives an upper limit on  $\alpha_3$ of 0.12 over the
entire allowed range of c when $b/\tau$ unification is imposed.
Possible
ways for  the determination of both c and $c'$ were discussed.Specifically
it is proposed that  a breakdown of $m_{\tilde{W}_1}-m_{\tilde{Z}_1}$
scaling relation[15],
 can help determine  $c'$.

This research was supported in parts by NSF grant number PHY--19306906.

\narrowtext

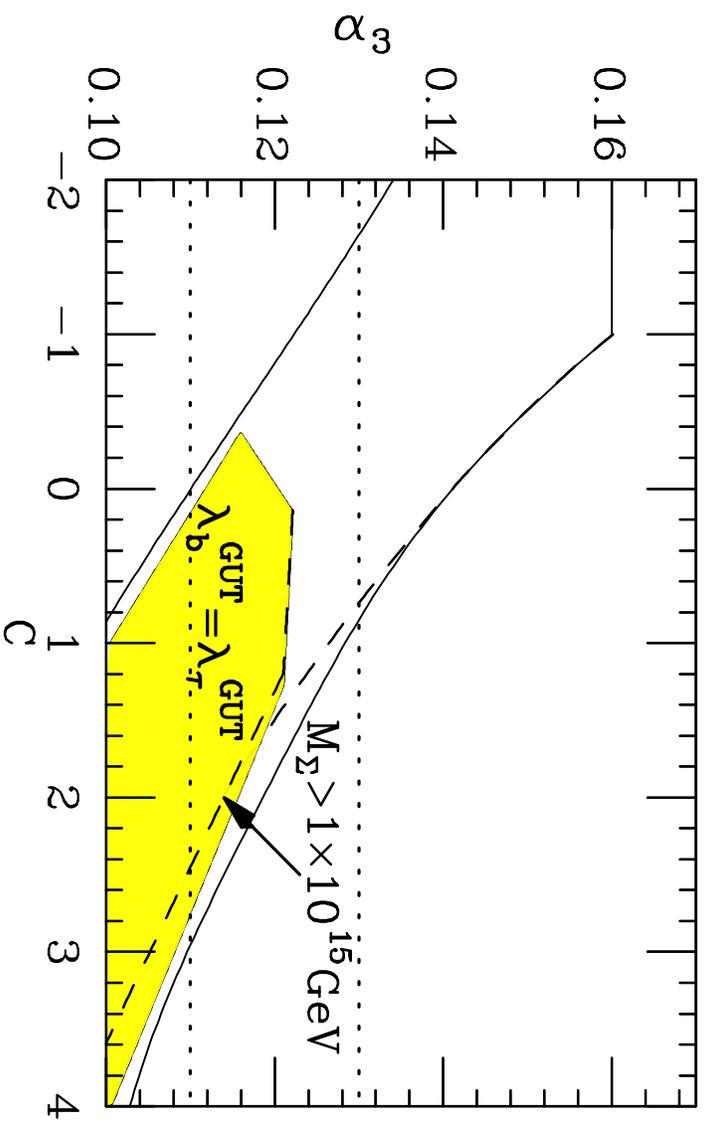
\begin{figure}
\caption{Allowed ranges of $\alpha_3$ as a function of c for $m_b=4.8GeV,
M_t=174\pm 16 GeV$ and $M_{SUSY}<10TeV$ and M=$10^{17}/(15)^{1/2}$.
The dotted lines show the bounds
on c coming from the LEP data($\alpha_3=0.12\pm 0.10$).Shaded(unshaded)
areas are the regions allowed by R.G. analyses with(without) $m_b/m_{\tau}$
constraint.}
\label{reduced}
\end{figure}

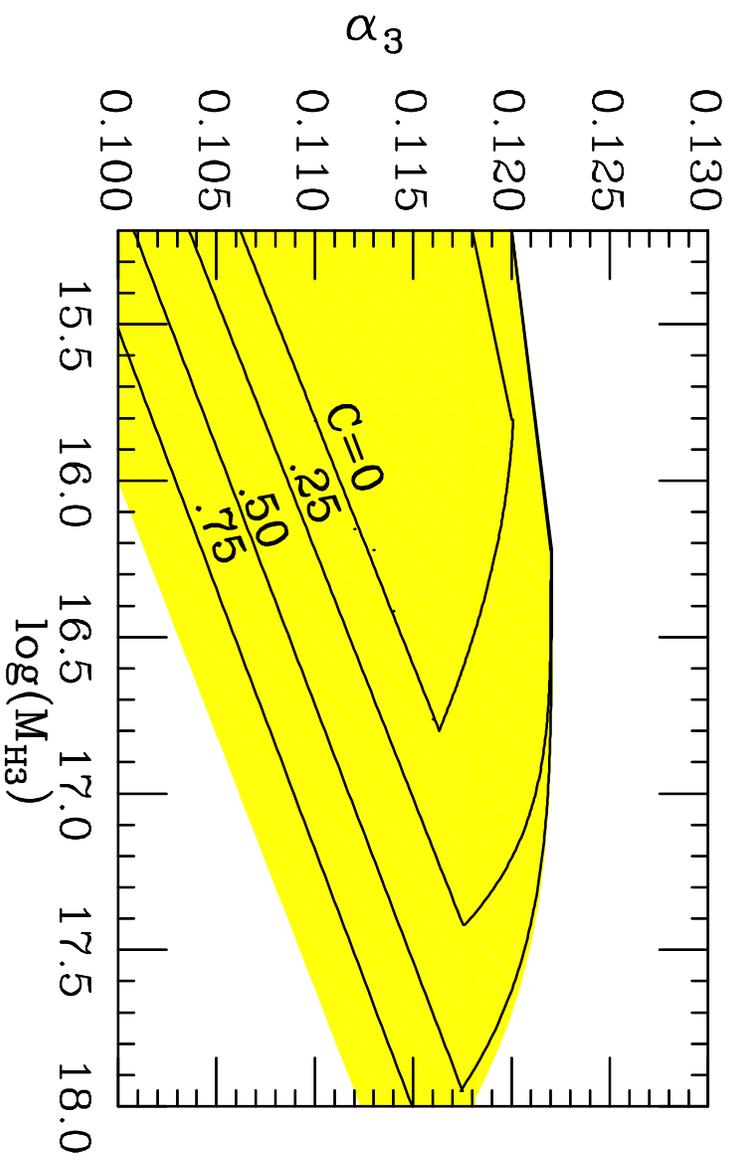
\begin{figure}
\caption{Allowed regions of $\alpha_3$ and $M_{H_3}$ for different values
of c.The upper limit of $\alpha_3$ is found to be insensitive to c due
to the strong $m_b/m_{\tau}$ constraint while the lower limit is not
strongly affected  by the constraint.As anticipated the fig. exhibits
scaling  with c of
the upper bound on $M_{H_3}$ as well as of the size of the allowed allowed
regions(The shaded area corresponds to $-0.4<c<1.0$).
 }
\end{figure}

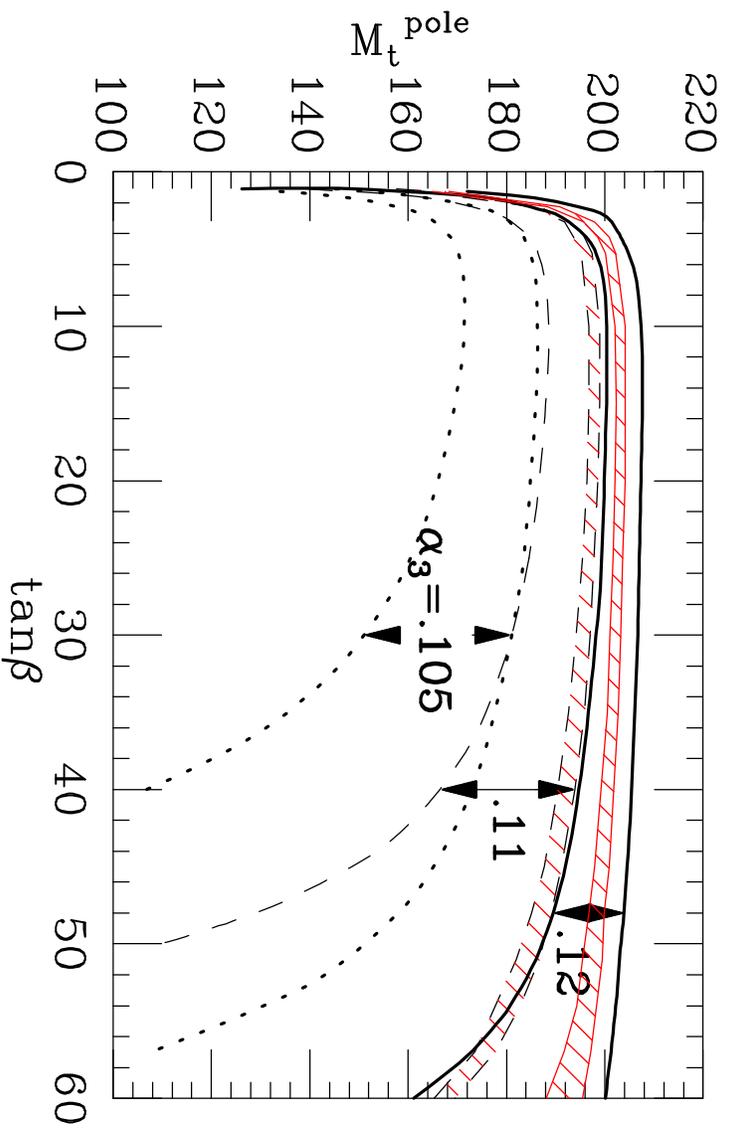
\begin{figure}
\caption{Allowed range of $M_t^{pole}$ as a function of $tan\beta$ for
$\alpha_3=0.105$(dot), $\alpha_3=0.11$(dash) and $\alpha_3=0.12$(solid)
($m_b=4.8 {\rm GeV  \& M_{SUSY}}< 10 {\rm TeV}$).The shaded areas correspond to
$c=0$.
We find that
$c<0.9$ is not allowed for $\alpha_3=0.105$ for $M_{SUSY}<10$TeV.
Within each region of
fixed $\alpha_3$ one finds that  curves for $c>0$($c<0$) always
lie higher(lower)than those
for the c=0 case.}
\end{figure}
%\end{multicols}
\end{document}